\documentclass[manuscript,screen]{acmart}

\AtBeginDocument{%
  \providecommand\BibTeX{{%
    \normalfont B\kern-0.5em{\scshape i\kern-0.25em b}\kern-0.8em\TeX}}}

\setcopyright{acmlicensed}
\copyrightyear{2023}
\acmYear{2024}
\acmDOI{XXXXXXX.XXXXXXX}

\acmConference[ICMHI '24]{Comparative Study of Generative Models for Early Detection of Failures in Medical Devices}{May 17--19,
  2024}{Yokohama, Japan}
\acmISBN{978-1-4503-XXXX-X/18/06}

\begin{document}

\title{Comparative Study of Generative Models for Early Detection of Failures in Medical Devices}

\author{Binesh Sadanandan}
\authornote{All authors contributed equally to this research.}
\email{bsada1@unh.newhaven.edu}
\orcid{}
\author{Bahareh Arghavani Nobar}
\authornotemark[1]
\email{bnobar@unh.newhaven.edu}
\author{Vahid Behzadan}
\authornotemark[1]
\email{vbehzadan@newhaven.edu}
\affiliation{%
  \institution{Department of Electrical and Computer Engineering and Computer Science, University of New Haven}
  \streetaddress{300 Boston Post Rd}
  \city{West Haven}
  \state{CT}
  \country{USA}
  \postcode{06516}
}

\renewcommand{\shortauthors}{Sadanandan and Nobar et.al}

\begin{abstract}
The medical device industry has significantly advanced by integrating sophisticated electronics like microchips and field-programmable gate arrays (FPGAs) to enhance the safety and usability of life-saving devices. These complex electro-mechanical systems, however, introduce challenging failure modes that are not easily detectable with conventional methods. Effective fault detection and mitigation become vital as reliance on such electronics grows. This paper explores three generative machine learning-based approaches for fault detection in medical devices, leveraging sensor data from surgical staplers—a class 2 medical device. Historically considered low-risk, these devices have recently been linked to an increasing number of injuries and fatalities. The study evaluates the performance and data requirements of these machine-learning approaches, highlighting their potential to enhance device safety.
\end{abstract}


\received{10 December 2023}

\maketitle

\section{Introduction}
The US Food and Drug Administration (FDA) defines a medical device as an instrument or apparatus intended to diagnose, cure, mitigate, treat, or prevent human diseases. The medical device industry is leaping forward by relying on electronics to improve the safety and performance of life-saving medical devices. The complex electronics, including microchips and FPGAs, run powerful software that helps further improve usability. However, the complex electro-mechanical systems introduced a new set of failure modes that are often difficult to identify and mitigate through traditional test protocols. Between 2006 and 2011, 5,294 recalls and 1,154,451 adverse events from device failures resulted in 92,600 patient injuries and 25,800 deaths  \cite{alemzadeh}. The Food and Drugs Administration (FDA) data also reveals that one in every three medical devices that use software for operation has been recalled due to failures in the software itself. A study in 2017 \cite{ronquillo2017software} found that a total of 627 software devices (1.4 million units) were subject to recalls, with 12 of these devices (190,596 units) subject to the highest risk recall level by the FDA. At least one equipment failure was noted in 38.8\% of operative procedures, whereas laparoscopic procedures had a high 41.9\%,increasing the total duration of the procedure by 7\%. Of the incidences, 19\% led to severe complications for the patient but with no morbidity or mortality \cite{courdier2009equipment}. Prolonged operative duration significantly increases the cost and the likelihood of complications, and decreased operative time is recommended as a universal goal for surgeons. To offset the cost of complex medical devices, manufacturers design and manufacture multi-use equipment that must undergo rigorous reprocessing between their use to maintain sterility. While this meets the objective of cost reduction, it also introduces additional failure modes that vary from hospitals\cite{di2019survey}. The evolution of technology in connectivity and data collection paired with sensors opens the door for preventive and predictive maintenance to mitigate failure in critical devices. The predictive and preventive maintenance strategy uses fault detection techniques leveraging data, signal, process, or knowledge-based methods. These techniques detect and prevent faults that otherwise would fail, causing a safety issue or degraded performance. Medical devices vary in their technological and engineering sophistication based on their intended use; however, in many cases, human life depends on their performance and reliability. 

For the scope of this analysis, we define \textit{fault} as an unpermitted deviation of at least one characteristic property or parameter from the expected normal condition in a piece of equipment or a system that may lead to failure. Anomaly detection refers to the problem of finding patterns in data that do not conform to expected behavior. Fault detection is the monitoring approach to detecting, isolating, and identifying faults using the concept of redundancy, either hardware redundancy or analytical redundancy. A search in the IEEE (Institute of Electrical and Electronics Engineers) Explore digital library for the keywords "Fault Detection" and  "Fault prevention" returned more than 4000 published papers between 2012 and 2022. These papers were proceedings from various conferences like the Renewable Engineering Conference (IREC)  and  IEEE Conference on Systems, Process, and Control (ICSPC), which serve as a good indicator of research interests in fault prevention. A recent literature survey on failure prognosis and applications by Kordestani et al. \cite{kordestani2019failure} reviewed the state-of-the-art techniques in fault detection. However, it lacked focus on medical devices where health data privacy is as crucial as its reliability. A reactive bottom-up Techno-centrism approach for security and privacy requirements is inadequate to protect patient privacy \cite{abouelmehdi2017big} and does not satisfy the regulatory requirement for medical devices. 

The main contributions of this paper are as follows:

    \begin{itemize}
        \item To the best of our knowledge, we are the first to evaluate the application of Generative methods to detect faults in medical devices.
        \item Our study compares and discerns the three generative algorithms for fault detection, accuracy, learning methods, and data requirements.
        \item Our evaluation used a time-series real-world dataset generated from a sensor embedded in a class 2  medical device, informing about the applicability of  our study across a wide variety of medical devices
    \end{itemize}
    
\section{Related Work}
Fault detection is a mature field of research with established approaches \cite {maktoubian2019iot}; however, the recent advancements in connectivity in medical devices have opened the door for data-driven and context-aware fault detection algorithms. Alemzadeh et al.\cite{alemzadeh} analyzed various safety-critical failures reported in Medical devices based on the US Food and Drug Administration(FDA) database. The scope of this study was to investigate the root causes of computer-based medical device failures and their effects; their study results accentuated the need for robust error-detection techniques with validated fail-safe mechanisms. Maktoubian et al. l \cite{maktoubian2019iot} reviewed the maintenance strategy for medical devices based on an Internet of Things(IoT) enabled framework that combined a monitoring and analytic module. The authors proposed an analytic module leveraging machine learning and predictive methods. Guo et al. Exploring the landscape of fault detection in medical devices further, we encounter various innovative approaches. Notably, DEMATEL \cite{guo2020research}  emerged as a method with heightened development efforts and specific applicability constraints. In this context, Meneghetti et al. \cite{meneghetti2018fault} made significant strides by developing an unsupervised approach for fault detection in artificial pancreas systems, utilizing outlier factor approximation, a method based on observation density for anomaly detection. To our knowledge, our research is the first to review Generative Machine Learning approaches and describe the learning, tuning, and selection based on a real-world medical device.

\section{Dataset Description}

We utilized two time-series univariate datasets for bench-marking from two distinct sources in this study. (i) DS1 - dataset consists of vibration measurements from actual helicopter tests provided by Airbus (ii) DS2 - dataset consists of motor current drawn from a laparoscopic surgical stapler generated in lab tests. 
\begin{itemize}

 \item DS1: The Airbus helicopter dataset, collected and released by Airbus SAS \cite{doi:10.1177/1748006X21994446}, serves as a valuable benchmark for studying vibration measurements in helicopter flight tests by placing accelerometers at different positions on the helicopter to measure the vibration in all operating conditions. The dataset consists of multiple one-dimensional time series captured at a constant frequency of 1 kHz. It is divided into two main parts: a training set and a test set. The training set comprises 1677 sequences of regular flights, providing a comprehensive representation of the expected vibration patterns during regular helicopter operations. These sequences serve as a valuable resource for learning the expected behavior of accelerometer data and training fault detection algorithms. The test set consists of 594 sequences, including normal and abnormal flights, enabling thorough evaluation and validation of anomaly detection methods.

\item DS2: This dataset, from a leading manufacturer, includes motor current readings from lab-simulated tests on a Powered Laparoscopic Surgical Stapler. These staplers, used in gastrointestinal, gynecologic, and thoracic surgeries, deliver staples for tissue connection and healing \cite{geller2022food}. The dataset contains varied-length one-dimensional time series signals recorded at 1 kHz, reflecting the different operational durations in surgeries. This structure, common in temporal data collection (e.g., sensor recordings, audio waveforms), enables extensive analysis and modeling to identify inherent patterns. Anomalies were artificially introduced into the dataset, marked by change points, and verified with a high-speed camera. This process, while ensuring data confidentiality through value alteration, aids in distinguishing between normal and anomalous events, enhancing fault detection in the stapler dataset.

\end{itemize}
By incorporating this dataset alongside the Airbus SAS dataset, which contains vibration measurements from accurate helicopter flight tests, we aim to advance fault detection techniques in surgical devices. The combination of these datasets allows for a comprehensive analysis of anomaly detection methods in different domains, providing insights into the applicability and effectiveness of approaches such as hidden Markov models (HMM), generative adversarial networks (GAN), and Variational Auto-Encoders (VAE). By leveraging the unique characteristics of each dataset, we can enhance the understanding of fault detection algorithms and improve safety and reliability in surgical procedures.

\section{Methodology}

\subsection{Data Pre-processing} \label{clustValid}
The preprocessing phase is critical in preparing the Airbus and surgical stapler datasets for analysis. Its effectiveness significantly impacts the quality of subsequent anomaly detection results. This section dives into the specific preprocessing steps employed for each dataset. For the surgical stapler dataset, we divided the dataset into separate training and testing subsets based on their labels. The training subset includes instances labeled as 0, while the testing subset consists of instances marked as 1. 
The following key steps are involved in data processing:
    \begin{itemize}
            \item Sequence Padding: As a part of our data preprocessing pipeline, we applied sequence padding to ensure uniform input lengths, a fundamental requirement for sequence-based models. This involves computing the maximum sequence length based on the 'TimeMS' column grouped by 'id' and subsequently padding the sequences using the \texttt{pad\_sequences} function. The goal is to standardize sequence lengths by pre-padding shorter sequences with zeros. The outcomes are consolidated into a new DataFrame named \texttt{padded\_df}, which includes the selected features, 'id', 'Padded\_TimeMS', and 'Anomaly\_At\_Time\_ms.' This structured \texttt{padded\_df} is the foundation for subsequent preprocessing steps and model training, promoting compatibility with sequence-based anomaly detection algorithms.
        \item Windowing: The initial step involves segmenting the continuous time series data into fixed-size windows. This segmentation approach is instrumental in enabling localized analysis and extracting underlying patterns. Each window encapsulates a distinct time frame, ensuring that patterns within specific intervals are adequately captured.
        \item Feature Extraction: Statistical features are computed within each window to extract meaningful insights. Four critical features were calculated: mean, median, skewness, and kurtosis. The mean measures central tendency, while kurtosis captures the distribution's tail behavior. These features collectively contribute to capturing distinctive data characteristics, aiding in anomaly differentiation.
        \item Data Splitting: With our data appropriately labeled, we proceed to create distinct training and testing sets in the third step. The training set is composed of normal instances, accounting for 80\% of the normal samples. This subset ensures that the model is exposed to a majority of normal data during the training phase. The remaining normal samples, along with all faulty samples, constitute the testing set. This arrangement guarantees that the testing set comprises both normal and faulty instances, thereby facilitating comprehensive model evaluation.
        \item Data Scaling: For effective model training and convergence, data scaling is implemented in the fourth step. This involves standardization using a \texttt{StandardScaler}, which is instantiated and fitted to the training data. The scaler computes the mean and standard deviation of the training data to ensure consistent scaling. Subsequently, both the training and testing feature data are transformed using the fitted scaler. This transformation centers the data around zero and scales it to have a standard deviation of one, contributing to stable and efficient model training.
    \end{itemize}
\subsection{Training} \label{clustValid}
\subsubsection{\textbf{HMM Training Procedure}}

In our HMM training procedure, we opted for two hidden states based on the nature of our datasets and the specific requirements of our anomaly detection task. These states are intended to represent the 'normal' and 'anomalous' conditions within the data. The choice was also influenced by the simplicity and computational efficiency that a two-state model offers, making it well-suited for real-time anomaly detection. This binary state approach aligns with the typical operational states of the medical devices under study, where they are either functioning as expected ('normal') or exhibiting fault characteristics ('anomalous'). This model structure is not only computationally efficient but also effectively captures the transition dynamics between normal and faulty conditions, which is crucial for early fault detection in medical devices.
The determination of the discriminative threshold for the HMM was a crucial aspect of our anomaly detection process. We employed a data-driven approach to establish this threshold. Specifically, we calculated the reconstruction error across the training dataset, which consisted solely of normal instances, and then analyzed the distribution of these errors. The threshold was set at a value that maximized the separation between normal and anomalous states, which we determined through statistical analysis and cross-validation. This threshold was further refined by considering the false positive rate we were willing to tolerate, ensuring that the model remained sensitive to anomalies while minimizing the incidence of false alarms. This methodical approach ensured that the threshold was tailored to the specific characteristics of our datasets, thereby enhancing the model's ability to distinguish between normal and anomalous conditions accurately.
After the training phase, the Hidden Markov Model (HMM) is used to predict the sequence of hidden states in a combined dataset of faulty and normal instances. We used the reconstruction error to differentiate between the actual anomaly and the predicted hidden states by the HMM. 

\subsubsection {\textbf{VAE Training Procedure}}

The VAE architecture encompasses an encoder and a decoder. The encoder, responsible for compressing the input data into a lower-dimensional representation, involves two fully connected layers. These layers are activated using the Rectified Linear Unit (ReLU) activation function. This activation function ensures that only positive activations are propagated, introducing non-linearity into the mapping process. Notably, both layers employ L2 regularization, which constrains the network weights and helps mitigate overfitting. This regularization technique adds a penalty term to the loss function based on the squared magnitude of weights.On the other hand, the decoder operates inversely, aiming to reconstruct the input data from the latent space. This phase employs two additional fully connected layers. The activation function utilized here is the sigmoid activation, which ensures the output remains within the $[0, 1]$ range. Similarly, L2 regularization is applied to the weights of these decoder layers. The training loss function is a sum of two constituents: a reconstruction loss and a Kullback-Leibler (KL) divergence loss. The reconstruction loss gauges the dissimilarity between the decoded output and the original input. The KL divergence loss enforces the learned latent distribution to mirror a standard Gaussian distribution. The VAE learns to generate meaningful latent representations by optimizing this composite loss function while ensuring the reconstructed output closely aligns with the input data. The optimization process employs the Adam optimizer with a specific learning rate, facilitating the convergence of the model. Through iterative optimization over 300 epochs, the VAE refines its encoder and decoder parameters, ultimately achieving enhanced data reconstruction and latent space organization.

\subsubsection {\textbf{GAN Training Procedure}}

 The GAN architecture consists of a generator and a discriminator. Prior to input, the data, which may be in the form of images or data points, is flattened into 1D vectors to align with the neural network input format. In the generator of our GAN model, the densely connected layers are configured with specific dimensions to optimize the generation process. The first layer has 256 units, followed by a second layer with 512 units, leading to the output layer that matches the dimensionality of the flattened input. Each layer uses the LeakyReLU activation function for enhanced gradient flow during training. The discriminator, conversely, consists of layers with a reverse configuration: starting with 512 units and then reducing to 256 units in the next layer. This structure allows the discriminator to effectively process and classify the incoming data. Both the generator and discriminator layers are interconnected, ensuring a balance in the model’s capacity to generate and discern data.In the discriminator, we employ dropout regularization, with a dropout rate of 0.4, to prevent overfitting and enhance the model's generalization capabilities. Dropout regularization works by randomly 'dropping out' a subset of neurons during the training phase, forcing the network to learn more robust features that are not reliant on any specific set of neurons. This technique is particularly effective in the discriminator as it ensures that the model does not overly rely on certain features, thereby maintaining its effectiveness in classifying real and synthetic data across various iterations. For the optimization of both the generator and discriminator, we use the Adam optimizer, known for its efficiency in handling sparse gradients and adaptive learning rate capabilities. We set the initial learning rate to 0.0002 with first and second-moment decay rates (beta1 and beta2) of 0.5 and 0.999, respectively. These hyperparameters were chosen to balance fast convergence and stability during training. The learning rate was kept relatively low to allow for finer adjustments in the weights, leading to a better generation of synthetic data that closely resembles the real data distribution.
 
\begin{table*} 
    \centering
    \caption{Comparison of Generative Methods for Fault Detection}
    \label{table:compare_algo}  
    \footnotesize 
    \begin{tabular}{|p{2.5cm}|p{3.5cm}|p{3.5cm}|p{3.5cm}|}
    \hline
    \multicolumn{4}{|c|}{\textbf{Generative Models - Fault/Anomaly Detection}} \\
    \hline
    \textbf{Algorithm} & \textbf{VAE} & \textbf{GAN} & \textbf{HMM} \\
    \hline
    \textbf{Architecture} & Convolutional AE & Convolutional & GMM \\
    \hline
    \textbf{Benefit} & Ability to work with higher dimensional input along with regularization on latent space helps VAE learn the input distribution effectively & GANs are powerful sample generators that can learn the data mapping scheme to determine anomaly scores & Double Embedded Stochastic process with two hierarchy levels, capable of learning complex Stochastic processes \\
    \hline
    \textbf{Shortcomings} & Training VAEs using ELBO can lead to a suboptimal generative model, biased towards ones with simpler posteriors & Requires optimization steps for each new input; potential for poor test-time performance and difficulty interpreting anomaly scores. The risk of non-convergence & Complex pre-processing steps require sequential data comparability. Statistical analysis needs a large data volume \\
    \hline
    \textbf{Type of Data} & Image; with Sequence to Sequence architecture, can model text, voice, time series & Image/data generation; variants developed for sequential data & Sequential data \\
    \hline
    \textbf{Learning Approach} & Maximum Likelihood with Explicit Density (Variational Approximation) & Maximum Likelihood with Implicit Density & Maximum Likelihood with Explicit Density (Markov Chain) \\
    \hline
    \textbf{Objective} & Inference by matching latent distribution to the original data distribution & Learn the distribution of the original data & Model unobservable hidden behaviors using observable data \\
    \hline
    \textbf{Performance Metrics} & Log Likelihood and Error & Accuracy and Error & Model parameters estimated using the Baum-Welch algorithm \\
    \hline
    \end{tabular}
\end{table*}

\subsection{Performance Metrics}
To evaluate various anomaly detection algorithms and their applicability in Medical devices, we bench-marked each of them on their, \textbf{Accuracy}, which quantifies the fraction of predictions our model accurately classified. It measures the overall correctness of the model's predictions,\textbf{Precision}, reflects the proportion of instances classified as positive that were actually correct. It gauges the model's ability to avoid false positive predictions and \textbf{Recall} also known as sensitivity or true positive rate, quantifies the proportion of true positive classifications that were identified correctly. It indicates the model's effectiveness in capturing positive instances.
    
The effectiveness of the HMM, GAN, and VAE models in detecting anomalies in both the Airbus and Stapler datasets is quantitatively summarized in Table \ref{table:combined_metrics}, where we present a comparative analysis of their accuracy, precision, and recall metrics. This table provides a clear overview of each model's performance, highlighting their strengths and areas of improvement in the context of fault detection in medical devices.

\begin{table}[h]
    \centering
    \caption{Evaluation Metrics on Airbus and Stapler Datasets}
    \label{table:combined_metrics}  
    \begin{tabular}{|p{2cm}||p{1.5cm}|p{1.5cm}|p{1.5cm}||p{1.5cm}|p{1.5cm}|p{1.5cm}|}
    \hline
    \multicolumn{1}{|c||}{Model} & \multicolumn{3}{c||}{Airbus Dataset} & \multicolumn{3}{c|}{Stapler Dataset} \\
    \hline
    & Accuracy & Precision & Recall & Accuracy & Precision & Recall \\
    \hline
    HMM & 97\% & 100\% & 97\% & 82\% & 100\% & 82\% \\
    GAN & 94\% & 100\% & 94\% & 95\% & 100\% & 95\% \\
    VAE & 97\% & 98\% & 97\% & 97\% & 100\% & 97\% \\
    \hline
    \end{tabular}
\end{table}

\subsection{Conclusion}

Our analysis of the stapler dataset has led to some significant findings. In our comprehensive assessment of generative models for fault detection, as detailed in Table \ref{table:compare_algo}, we highlight the distinct characteristics and performance metrics of VAE, GAN, and HMM. This comparative analysis further substantiates our findings, emphasizing the strengths and limitations of each model in the context of anomaly detection in medical devices. The superior performance of the VAE and GAN models compared to the HMM can be attributed to their inherent architectural advantages. VAE's ability to generate a compact and continuous latent space representation of the data, and GAN's proficiency in capturing complex data distributions provide a more nuanced understanding of 'normal' vs. 'anomalous' patterns. This is especially beneficial in dealing with medical device data, which often exhibits subtle and complex variations. In contrast, the HMM, while effective, is somewhat limited by its reliance on the assumption of Markovian dynamics and may not capture the intricacies of the data as effectively as the neural network-based approaches of VAE and GAN.

Contrary to our initial assumption that a window size of 1024 milliseconds was optimal, model performance indicates that increasing this to 2048 milliseconds enhances performance. This modification highlights the critical role of experimenting with different parameters in anomaly detection tasks. Our convolutional-based models, particularly the GAN and VAE variants, have surpassed the HMM, achieving an 82\% accuracy rate. This success demonstrates our model's proficiency in accurately identifying anomalies within the stapler dataset. Remarkably, these results were achieved without resorting to windowing techniques, further underscoring the inherent strength of our model. These specific results not only underscore the efficacy of VAE and GAN in fault detection but also pave the way for understanding the broader implications of our findings. The practicality of a reconstruction-based approach, as highlighted by our analysis, signals a paradigm shift in anomaly detection strategies for medical devices.

Our findings indicate that Generative Methods, particularly with a window size of 2048 milliseconds, are highly effective in anomaly detection. The superior performance of our GAN and VAE models highlights their potential for wide-ranging applications in real-world scenarios. This underscores the importance of parameter tuning and feature selection for optimizing anomaly detection systems. Through the Airbus dataset, we gained valuable insights into the strengths of various anomaly detection models. Our evaluation, which focused on accuracy, precision, and recall, provided a comprehensive understanding of model performance.

The VAE exhibited outstanding results, achieving a remarkable accuracy rate of 97\%. Following closely, the GAN demonstrated a commendable performance with an accuracy rate of 94\%. Significantly, these results surpass the findings of previous studies,\cite{chen2021deep}, VAE achieved an accuracy of 90\%, and GAN achieved an accuracy of 86\% on the Airbus dataset. Therefore, our implementation has achieved superior results for these two models. Finally, the Hidden Markov Model (HMM) showcased respectable performance with an accuracy rate of 91\%. This result is also noteworthy compared to the HMM accuracy of 95\% reported in the study~\cite{juan2021automatic} for the Airbus dataset. Our comparison results highlight the different levels of effectiveness across the models, with VAE and GAN emerging as the most successful in our tests. Additionally, our study concentrated on three generative, machine learning-based methods for detecting faults in time-series data. We emphasized their operational performance and the data required for training. Our main aim was to assess the practicality of a reconstruction-based approach for fault detection, a method broadly applicable to devices producing time-series data. Our analysis covered the training phase, where the model learns the normal data distribution, and the detection phase, where this knowledge is applied to identify anomalies. Although our techniques have shown high accuracy in detecting anomalies, we recognize the need for a more extensive analysis and the potential for further improvements. Future research should explore the hardware requirements for deploying these methods on FPGAs and Edge-based devices.

\section{Data Availability and Code Repository}
\begin{itemize}
    \item \textbf{Airbus Helicopter Accelerometer Dataset:} This dataset is publicly available and can be accessed from \href{https://www.research-collection.ethz.ch/handle/20.500.11850/415151}{ETH Zürich's Research Collection}.
    \item \textbf{Surgical Stapler Dataset:} Due to security and privacy considerations, the Surgical Stapler dataset is not publicly available.
\end{itemize}
The code developed for this study is available in a GitHub repository and can be accessed via the following link: \href{https://github.com/UNHSAILLab/Fault-Detection-on-Surgical-Stapler-}{GitHub Repository}.

\begin{acks}
We would like to extend our heartfelt gratitude to Medtronic's Surgical Operating Unit for granting us access to the data essential for this experiment. Their support has been invaluable in the advancement of this research. It is important to note that the data obtained from the medical device were sourced from a lab experiment and are not clinical data. Furthermore, all real values in the experiment have been anonymized for data protection and privacy. 
\end{acks}

\bibliographystyle{ACM-Reference-Format}
\bibliography{Fault_Detection_Paper}


\begin{thebibliography}{13}


\ifx \showCODEN    \undefined \def \showCODEN     #1{\unskip}     \fi
\ifx \showDOI      \undefined \def \showDOI       #1{#1}\fi
\ifx \showISBNx    \undefined \def \showISBNx     #1{\unskip}     \fi
\ifx \showISBNxiii \undefined \def \showISBNxiii  #1{\unskip}     \fi
\ifx \showISSN     \undefined \def \showISSN      #1{\unskip}     \fi
\ifx \showLCCN     \undefined \def \showLCCN      #1{\unskip}     \fi
\ifx \shownote     \undefined \def \shownote      #1{#1}          \fi
\ifx \showarticletitle \undefined \def \showarticletitle #1{#1}   \fi
\ifx \showURL      \undefined \def \showURL       {\relax}        \fi
\providecommand\bibfield[2]{#2}
\providecommand\bibinfo[2]{#2}
\providecommand\natexlab[1]{#1}
\providecommand\showeprint[2][]{arXiv:#2}

\bibitem[Abouelmehdi et~al\mbox{.}(2017)]%
        {abouelmehdi2017big}
\bibfield{author}{\bibinfo{person}{Karim Abouelmehdi}, \bibinfo{person}{Abderrahim Beni-Hssane}, \bibinfo{person}{Hayat Khaloufi}, {and} \bibinfo{person}{Mostafa Saadi}.} \bibinfo{year}{2017}\natexlab{}.
\newblock \showarticletitle{Big data security and privacy in healthcare: a review}.
\newblock \bibinfo{journal}{\emph{Procedia Computer Science}}  \bibinfo{volume}{113} (\bibinfo{year}{2017}), \bibinfo{pages}{73--80}.
\newblock


\bibitem[Alemzadeh et~al\mbox{.}(2013)]%
        {alemzadeh}
\bibfield{author}{\bibinfo{person}{Homa Alemzadeh}, \bibinfo{person}{Ravishankar~K Iyer}, \bibinfo{person}{Zbigniew Kalbarczyk}, {and} \bibinfo{person}{Jai Raman}.} \bibinfo{year}{2013}\natexlab{}.
\newblock \showarticletitle{Analysis of safety-critical computer failures in medical devices}.
\newblock \bibinfo{journal}{\emph{IEEE Security \& Privacy}} \bibinfo{volume}{11}, \bibinfo{number}{4} (\bibinfo{year}{2013}), \bibinfo{pages}{14--26}.
\newblock


\bibitem[Chen et~al\mbox{.}(2021)]%
        {chen2021deep}
\bibfield{author}{\bibinfo{person}{Kaixuan Chen}, \bibinfo{person}{Dalin Zhang}, \bibinfo{person}{Lina Yao}, \bibinfo{person}{Bin Guo}, \bibinfo{person}{Zhiwen Yu}, {and} \bibinfo{person}{Yunhao Liu}.} \bibinfo{year}{2021}\natexlab{}.
\newblock \showarticletitle{Deep learning for sensor-based human activity recognition: Overview, challenges, and opportunities}.
\newblock \bibinfo{journal}{\emph{ACM Computing Surveys (CSUR)}} \bibinfo{volume}{54}, \bibinfo{number}{4} (\bibinfo{year}{2021}), \bibinfo{pages}{1--40}.
\newblock


\bibitem[Courdier et~al\mbox{.}(2009)]%
        {courdier2009equipment}
\bibfield{author}{\bibinfo{person}{S{\'e}bastien Courdier}, \bibinfo{person}{Olivier Garbin}, \bibinfo{person}{Michel Hummel}, \bibinfo{person}{V{\'e}ronique Thoma}, \bibinfo{person}{Elizabeth Ball}, \bibinfo{person}{Romain Favre}, {and} \bibinfo{person}{Arnaud Wattiez}.} \bibinfo{year}{2009}\natexlab{}.
\newblock \showarticletitle{Equipment failure: causes and consequences in endoscopic gynecologic surgery}.
\newblock \bibinfo{journal}{\emph{Journal of minimally invasive gynecology}} \bibinfo{volume}{16}, \bibinfo{number}{1} (\bibinfo{year}{2009}), \bibinfo{pages}{28--33}.
\newblock


\bibitem[Di~Mattia et~al\mbox{.}(2019)]%
        {di2019survey}
\bibfield{author}{\bibinfo{person}{Federico Di~Mattia}, \bibinfo{person}{Paolo Galeone}, \bibinfo{person}{Michele De~Simoni}, {and} \bibinfo{person}{Emanuele Ghelfi}.} \bibinfo{year}{2019}\natexlab{}.
\newblock \showarticletitle{A survey on gans for anomaly detection}.
\newblock \bibinfo{journal}{\emph{arXiv preprint arXiv:1906.11632}} (\bibinfo{year}{2019}).
\newblock


\bibitem[Garcia et~al\mbox{.}(0)]%
        {doi:10.1177/1748006X21994446}
\bibfield{author}{\bibinfo{person}{Gabriel~Rodriguez Garcia}, \bibinfo{person}{Gabriel Michau}, \bibinfo{person}{Mélanie Ducoffe}, \bibinfo{person}{Jayant~Sen Gupta}, {and} \bibinfo{person}{Olga Fink}.} \bibinfo{year}{0}\natexlab{}.
\newblock \showarticletitle{Temporal signals to images: Monitoring the condition of industrial assets with deep learning image processing algorithms}.
\newblock \bibinfo{journal}{\emph{Proceedings of the Institution of Mechanical Engineers, Part O: Journal of Risk and Reliability}} \bibinfo{volume}{0}, \bibinfo{number}{0} (\bibinfo{year}{0}), \bibinfo{pages}{1748006X21994446}.
\newblock
\urldef\tempurl%
\url{https://doi.org/10.1177/1748006X21994446}
\showDOI{\tempurl}
\showeprint{https://doi.org/10.1177/1748006X21994446}


\bibitem[Geller(2022)]%
        {geller2022food}
\bibfield{author}{\bibinfo{person}{Jay Geller}.} \bibinfo{year}{2022}\natexlab{}.
\newblock \showarticletitle{Food and Drug Administration Issue Final Order of Surgical Staples Reclassification}.
\newblock \bibinfo{journal}{\emph{Journal of Clinical Engineering}} \bibinfo{volume}{47}, \bibinfo{number}{1} (\bibinfo{year}{2022}), \bibinfo{pages}{3--5}.
\newblock


\bibitem[Guo et~al\mbox{.}(2020)]%
        {guo2020research}
\bibfield{author}{\bibinfo{person}{Xingru Guo}, \bibinfo{person}{Aijun Liu}, \bibinfo{person}{Xia Li}, {and} \bibinfo{person}{Yaxuan Xiao}.} \bibinfo{year}{2020}\natexlab{}.
\newblock \showarticletitle{research on the intelligent fault diagnosis of medical devices based on a DEMATEL-fuzzy concept Lattice}.
\newblock \bibinfo{journal}{\emph{International Journal of Fuzzy Systems}} \bibinfo{volume}{22}, \bibinfo{number}{7} (\bibinfo{year}{2020}), \bibinfo{pages}{2369--2384}.
\newblock


\bibitem[Juan et~al\mbox{.}(2021)]%
        {juan2021automatic}
\bibfield{author}{\bibinfo{person}{Zuluaga-Gomez Juan}, \bibinfo{person}{Karel Vesely}, \bibinfo{person}{Blatt Alexander}, \bibinfo{person}{Petr Motlicek}, \bibinfo{person}{Dietrich Klakow}, \bibinfo{person}{Allan Tart}, \bibinfo{person}{Igor Szoke}, \bibinfo{person}{Amrutha Prasad}, \bibinfo{person}{Seyyed~Saeed Sarfjoo}, \bibinfo{person}{Pavel Kolcarek}, {et~al\mbox{.}}} \bibinfo{year}{2021}\natexlab{}.
\newblock \showarticletitle{Automatic Call Sign Detection: Matching Air Surveillance Data with Air Traffic Spoken Communications}. In \bibinfo{booktitle}{\emph{Proceedings of the 8th OpenSky Symposium 2020}}, Vol.~\bibinfo{volume}{59}. MDPI.
\newblock


\bibitem[Kordestani et~al\mbox{.}(2019)]%
        {kordestani2019failure}
\bibfield{author}{\bibinfo{person}{Mojtaba Kordestani}, \bibinfo{person}{Mehrdad Saif}, \bibinfo{person}{Marcos~E Orchard}, \bibinfo{person}{Roozbeh Razavi-Far}, {and} \bibinfo{person}{Khashayar Khorasani}.} \bibinfo{year}{2019}\natexlab{}.
\newblock \showarticletitle{Failure prognosis and applications—A survey of recent literature}.
\newblock \bibinfo{journal}{\emph{IEEE transactions on reliability}} (\bibinfo{year}{2019}).
\newblock


\bibitem[Maktoubian and Ansari(2019)]%
        {maktoubian2019iot}
\bibfield{author}{\bibinfo{person}{Jamal Maktoubian} {and} \bibinfo{person}{Keyvan Ansari}.} \bibinfo{year}{2019}\natexlab{}.
\newblock \showarticletitle{An IoT architecture for preventive maintenance of medical devices in healthcare organizations}.
\newblock \bibinfo{journal}{\emph{Health and Technology}} \bibinfo{volume}{9}, \bibinfo{number}{3} (\bibinfo{year}{2019}), \bibinfo{pages}{233--243}.
\newblock


\bibitem[Meneghetti et~al\mbox{.}(2018)]%
        {meneghetti2018fault}
\bibfield{author}{\bibinfo{person}{Lorenzo Meneghetti}, \bibinfo{person}{Matteo Terzi}, \bibinfo{person}{Gian~Antonio Susto}, \bibinfo{person}{Simone Del~Favero}, {and} \bibinfo{person}{Claudio Cobelli}.} \bibinfo{year}{2018}\natexlab{}.
\newblock \showarticletitle{Fault detection in artificial pancreas: A model-free approach}. In \bibinfo{booktitle}{\emph{2018 IEEE Conference on Decision and Control (CDC)}}. IEEE, \bibinfo{pages}{303--308}.
\newblock


\bibitem[Ronquillo and Zuckerman(2017)]%
        {ronquillo2017software}
\bibfield{author}{\bibinfo{person}{Jay~G Ronquillo} {and} \bibinfo{person}{Diana~M Zuckerman}.} \bibinfo{year}{2017}\natexlab{}.
\newblock \showarticletitle{Software-related recalls of health information technology and other medical devices: Implications for FDA regulation of digital health}.
\newblock \bibinfo{journal}{\emph{The Milbank Quarterly}} \bibinfo{volume}{95}, \bibinfo{number}{3} (\bibinfo{year}{2017}), \bibinfo{pages}{535--553}.
\newblock


\end{thebibliography}

\end{document}